\RequirePackage{fix-cm}

\documentclass[twocolumn]{svjour3}          
\smartqed  
\usepackage{graphicx}
\usepackage[dvipsnames]{xcolor}

\usepackage{hyperref}
\usepackage{siunitx}

\newcommand{\aap}{A\&A}

\begin{document}

\title{The CUBES Instrument Model and Simulation Tools. 
}
\subtitle{Their role in the project Phase A study.}


\author{Matteo Genoni \and
        Marco Landoni \and
        Guido Cupani \and
        Mariagrazia Franchini \and
        Roberto Cirami \and
        Alessio Zanutta\and
        Chris Evans\and
        Paolo Di Marcantonio\and
        Stefano Cristiani\and
        Andrea Trost \and
        Sonia Zorba 
}


\institute{M. Genoni, A.Zanutta \at
              INAF - Osservatorio Astronomico di Brera\\
              via E. Bianchi 46, 23807, Merate (LC), Italy\\
              \email{matteo.genoni@inaf.it}  
              \\
              \\
              M. Landoni \at 
              INAF - Osservatorio Astronomico di Cagliari, via della Scienza 5 09047 Selargius (CA), Italy
              \\
              \\
              \and
            S.~Cristiani,  R.~Cirami, G.~Cupani, P.~Di Marcantonio, M.~Franchini, S.~Zorba \at
            INAF - Osservatorio Astronomico di Trieste, via G.B. Tiepolo 11, 34143, Trieste, Italy
            \\
            \and
            C.~Evans \at
            UK Astronomy Technology Centre, Royal Observatory, Blackford Hill, Edinburgh EH9 3HJ, UK\\
            \\
            \and
            A.~Trost \at
            Department of Physics, University of Trieste, Via A. Valerio 2, 34127 Trieste, Italy
            \\
            \\
}

\date{Received: date / Accepted: date}

\maketitle

\begin{abstract}
We present the simulation tools developed to aid the design phase of the Cassegrain U-Band Efficient Spectrograph (CUBES) for the Very Large Telescope (VLT), exploring aspects of the system design and evaluating the performance for different design configurations.
CUBES aims to be the `ultimate' ultraviolet (UV) instrument at the European Southern Observatory (ESO) in terms of throughput, with the goal to cover the bluest part of the spectrum accessible from the ground (\SIrange{300}{400}{nm}) with the highest possible efficiency. Here we introduce the End-to-End (E2E) and the Exposure Time Calculator (ETC) tools. The E2E simulator has been developed with different versions to meet the needs of different users, including a version that can be accessed for use by the broader scientific community using a Jupyter notebook.
The E2E tool was used by the system team to help define the Phase A baseline design of the instrument, as well as in scientific evaluation of a possible low-resolution mode. 
The ETC is a web-based tool through which the science community are able to test a range of science cases for CUBES, demonstrating its potential to push the limiting magnitude for the detection of specific UV-features, such as abundance estimates of beryllium in main-sequence stars.


\end{abstract}

\section{Introduction}
\label{intro}
The Cassegrain U-Band Efficient Spectrograph (CUBES) instrument is a new spectrograph in development for one of the Cassegrain foci of the Very Large Telescope (VLT) at the European Southern Observatory (ESO). The goal is to maximise the efficiency at a spectral resolving power ($R$) of greater than 20000, over the bluest part of the spectrum accessible from the ground
(\SIrange{300}{400}{nm}). Given its performance at short wavelengths, CUBES will provide an unprecedented blue eye on the sky alongside the Extremely Large Telescope (ELT) that is now under construction on Cerro Armazones \cite{Ref-Evans_2018}.

A tenfold improvement in sensitivity compared to existing instruments in the near-UV will open-up exciting new observations across a broad range of scientific topics, as highlighted by many of the contributions in this Special Issue. These range from searches of water in the asteroid belt
\cite{Ref-Opitom}, a broad range of applications in studies of stellar evolution and nucleosynthesis (ranging from light elements such as beryllium up to iron-peak and heavier elements, e.g. \cite{Ref-Ernandes_2020}), to extragalactic observations such as the contribution of galaxies and active galactic nuclei to the cosmic UV background \cite{Ref-Vanzella}.

The successful design, integration and commissioning of an instrument like CUBES requires a staged development (in phases). This allows trade-offs between requirements and possible hardware configurations to maximise the performance and arrive at a final layout that satisfies the scientific requirements and enables the desired range of science applications. In the context of the conceptual design phase (Phase~A) of CUBES we developed two software tools to aid both the technical and scientific teams, an End-to-End (E2E) simulator and an Exposure Time Calculator (ETC). 

Briefly, E2E instrument simulators are software systems that simulate the instrument behaviour during astronomical observations, from the flux distribution of the scientific sources through to the raw data produced as output by the instrument detector. The synthetic raw frames from the E2E simulations can also be used to aid the development of the Data Reduction Software (DRS), allowing first evaluation of the scientific goals of the project and to give feedback to the technical team during design phases and trade-off studies. 

ETCs allow astronomers to accurately plan the time required for potential observing programmes by predicting the signal-to-noise ratio (SNR) for a given exposure time. ETCs are quick and user-friendly tools to evaluate instrument performance for a range of scientific cases under study, making them a valuable complement to studies with the E2E simulator.

We developed and used these tools during the CUBES Phase~A design study to evaluate different possible configurations for the instrument, and to investigate the performance of a possible low-resolution mode. We then focused on particular science cases where CUBES will outperform existing instruments in the ground-UV to quantify its performance; many of these scientific applications are explored in the other contributions in this Special Issue.

The paper is organized as follows. In Sect.~2 we briefly review the overall design of CUBES while in Sect.~3 we present the instrument model and discuss its application for the trade-off analyses and performance evaluation in Sect.~4. We present the ETC tool and an example application in Sect~5, with concluding remarks in Sect.~6.

\section{The CUBES instrument}
\label{sec_CUBES_General}
CUBES will be mounted at the VLT Cassegrain focus and the Phase~A instrument baseline includes the following subsystems:
\begin{itemize}
    \item A fore-optics subsystem that includes an atmospheric dispersion corrector (ADC), the pre-slit and an Acquisition and Guiding (A\&G) subsystem,
    \item Two image slicers (to enable different resolutions),
    \item A fibrelink to UVES,
    \item A calibration unit,
    \item Two spectrograph arms: arm\,1 (blue-optimized) and arm\,2 (red-optimized), both equipped with transmission gratings with a high groove density, working at the first order, and with 9k\,$\times$\,9k scientific detectors.
\end{itemize}

\begin{figure*}[!t]
\centering
  \includegraphics[width=0.75\textwidth]{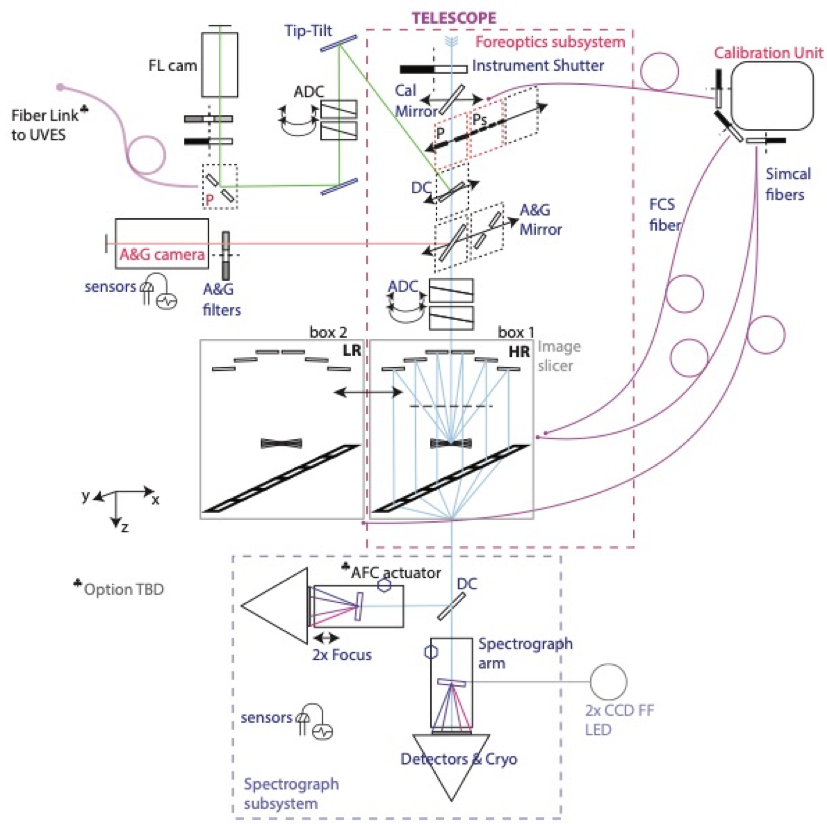}
\caption{CUBES functional overview. 
Taken from \cite{Ref-Zanutta_2021}.}
\label{fig:CUBES_functional_overview}       
\end{figure*}

The instrument design has two operational modes, either using CUBES alone or in combination with VLT-UVES \cite{Ref-UVES} for simultaneous observations via an optical fibrelink. A detailed description of the CUBES instrument is outside the scope of this paper and is given by \cite{Ref-Zanutta_2021}. Here we briefly outline its main characteristics and subsystems, as shown in the functional schematic of Fig.~\ref{fig:CUBES_functional_overview}.

\subsection{Fore-optics and A\&G subsystems}
The fore-optics subsystem provides a collimated beam for the ADC and the magnifying optics to feed the image slicer. The image slicer reformats the science object into narrower slits at the input to the spectrograph, enabling higher resolution with respect to a traditional slit. The Phase~A design includes two image slicers (for details see \cite{Ref-Zanutta_2021} and \cite{Ref-Calcines_2021}), to provide the option of high- and low-resolution modes (HR and LR, respectively).

The A\&G subsystem provides initial acquisition of the science target and secondary guiding capabilities during the exposure.

Finally, a dichroic mirror can be used to pick-off light to be relayed to the FibreLink subsystem for simultaneous observations with UVES.

\subsection{Fibrelink to UVES}
The FibreLink subsystem provides the option of relaying light to the UVES instrument on the Nasmyth platform for simultaneous observations at longer wavelengths ($\lambda$\,$>$\,420\,nm). The FibreLink subsystem includes an ADC and a fibre-guiding system.

\subsection{Calibration Unit}
The Calibration Unit provides light sources to perform calibration procedures, such as flat fielding, wavelength calibration, simultaneous wavelength calibration, alignment and flexure correction (in case an Active Flexure Compensation, AFC, System is required). It also provides calibration sources for UVES when used for simultaneous observations with CUBES.

\subsection{The spectrograph}
The spectrograph subsystem is composed of two arms. Each arm collimates the beam, disperses the light using a transmission grating (see \cite{Ref-Gratings_2021} for details) and refocuses the image with a camera on a 9k x 9k scientific detector. The possibility of including actuators for the AFC System (if required), has been foreseen from the beginning and will be carefully studied in next project phases.

\section{The CUBES E2E simulator}
\label{sec_E2E_sim}
\begin{figure*}
\centering
  \includegraphics[width=0.8\textwidth]{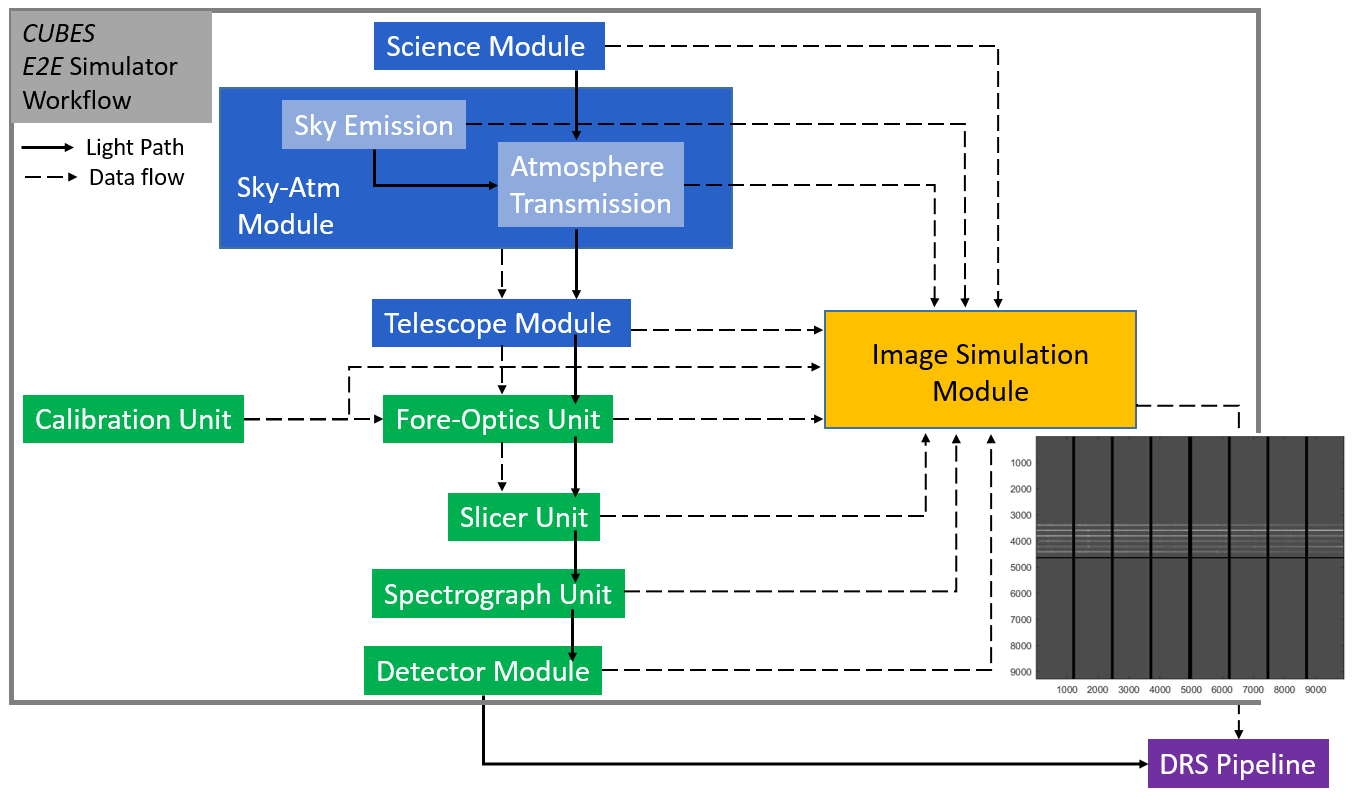}
\caption{E2E simulator workflow schematic description for the CUBES instrument (adapted from \cite{Ref-Genoni_2020}).
The solid arrows represent the light path, while the dashed arrows show the simulation data-flow, how the different modules and units are interfaced and their connection to the simulator core, which is the Image Simulation Module. The synthetic frame produced by the simulator is ingested by the data reduction system (DRS) pipeline. Green blocks are units of the Instrument Module, while blue blocks are related to simulation modules independent of the instrument.}
\label{fig:E2E_Tech_01}       
\end{figure*}

Following the work presented by \cite{Ref-Genoni_2020}, we adapted the existing E2E simulator architecture from other instruments and implemented the same modular approach for the CUBES simulator. Indeed, it is composed of different modules (each with specific tasks), units and interfaces, mimicking the functional architecture of the CUBES instrument, as described in the schematic workflow of Fig.~\ref{fig:E2E_Tech_01}. Further details of such a software architecture and its properties can be found in \cite{Ref-Genoni_2020}.

Three versions of the CUBES E2E have been developed for different user applications, featuring different capabilities (or functionalities) that were used during the Phase~A design study. 
These are:
\begin{itemize}
    \item \textit{Basic Version:} aimed at science users, to assess the performance of the instrument with respect to the science objectives, depending on the adopted instrument design.
    \item \textit{Parametric Version:} provides fast parametric simulations of the different possible design solutions and configurations, as an aid for the design development. 
    \item \textit{Full Version:} produces high-level simulated frames based on a detailed physical model of the instrument (including optical effects as the contribution of blurring of the point spread function (PSF), distortions and detector diffusion effects), to be used for the development and testing of the data reduction pipeline.
\end{itemize}

The \textit{Basic Version} of the simulator is implemented as a Python 3 library \cite{10.5555/1593511} and accessed through a Jupyter notebook \cite{jupyter}. The notebook includes both the basic code to perform the simulation and instructions on how to execute it\footnote{A compact version of the notebook, omitting all instructions, is also available.}. Users can freely configure the parameters of the simulation and run it. Advanced users may also inspect the library and possibly edit its functions through the Jupyter interface. The library and the notebook are available as a Git repository through GitHub\footnote{\url{https://github.com/gcupani/cubes}.} for download on local machines. It is also hosted on Binder \cite{binder}, an online tool to execute Jupyter notebooks on a remote server interactively\footnote{\url{https://bit.ly/cubes_e2e}. Note that the Binder version is limited in memory usage and is therefore not guaranteed to be executed with arbitrary choices of the simulation parameters.}.

The \textit{Full Version} is currently written in MATLAB and uses libraries for specific functionalities, as well as an ad-hoc software wrapper we developed to interface with other software, such as the commercial optical ray tracing Zemax-OpticStudio.
The \textit{Parametric Version} is also developed in MATLAB; it shares some modules and parts of code with the \textit{Full Version} (e.g. the Science Object Module), while it employs an ad-hoc parametric description of the spectrograph to run parametric simulations.

Given the different purposes and functionalities, the first version is available on-line for the scientific community, while the other two are used by the technical team.

\subsection{E2E Modules description and functionalities}
\label{subsec_E2E_Modules}
We now describe the module functionalities implemented for the specific needs of the CUBES instrument (building on the functionalities described by \cite{Ref-Genoni_2020}):

\paragraph{Science Object Module:} generates a synthetic 1D spectrum of an astronomical source for a user-defined set of parameters (e.g. flux distribution, magnitude), at a resolution higher than the selected instrument mode (i.e. HR or LR). A user-defined spectrum (in ASCII format) can also be uploaded as the input spectrum, where the user must ensure that the resolution is again greater than that of the selected HR or LR mode.

\paragraph{Sky radiance and atmospheric transmission:} spectra are modelled by the Sky-Atmosphere module invoking a dedicated library built using the ESO {\sc skycalc} tool (REFF-SkyCalc). These spectra are loaded from the library according to the adopted observing conditions for the simulation, i.e. moon phase, airmass and precipitable water vapour. 

\begin{figure*}
  \includegraphics[width=1\textwidth]{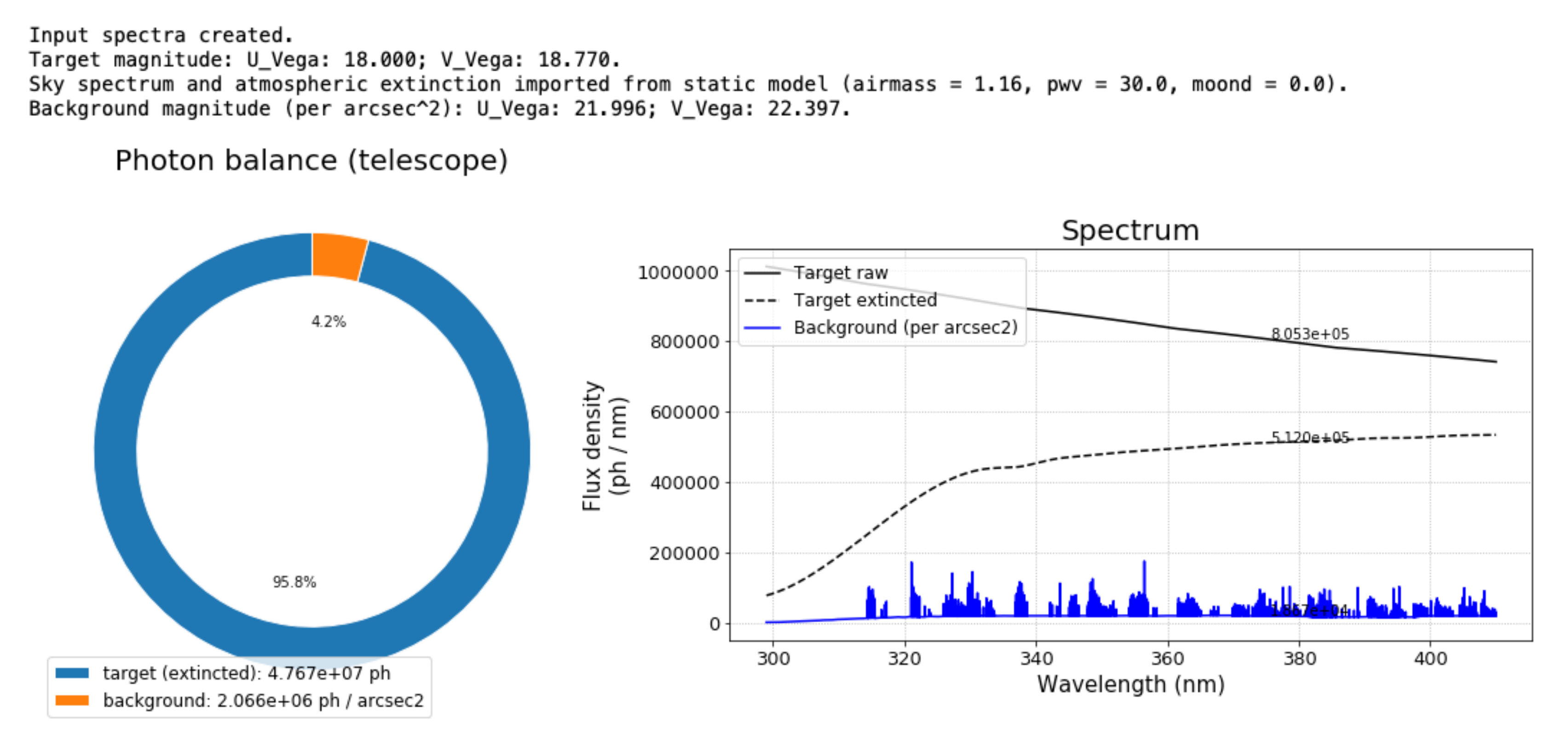}
\caption{Example of the input spectra in the \textit{Basic Version} E2E notebook (for a flat spectrum of $U$\,$=$\,18\,mag). The integrated flux from the target and the sky is computed assuming the collecting area of the primary mirror of a VLT unit telescope and a detector integration time of \SI{3600}{s}. \textit{Left:} photon balance between the target and the sky background. \textit{Right:} spectra of target (with extinction) and background.}
\label{fig:E2E_Sci_01_ObjSky}       
\end{figure*}

A typical output of these modules is shown in Fig.~\ref{fig:E2E_Sci_01_ObjSky}, including general information about the target and sky background, their photon balance, and their spectra.

\begin{figure}
  \includegraphics[width=0.5\textwidth]{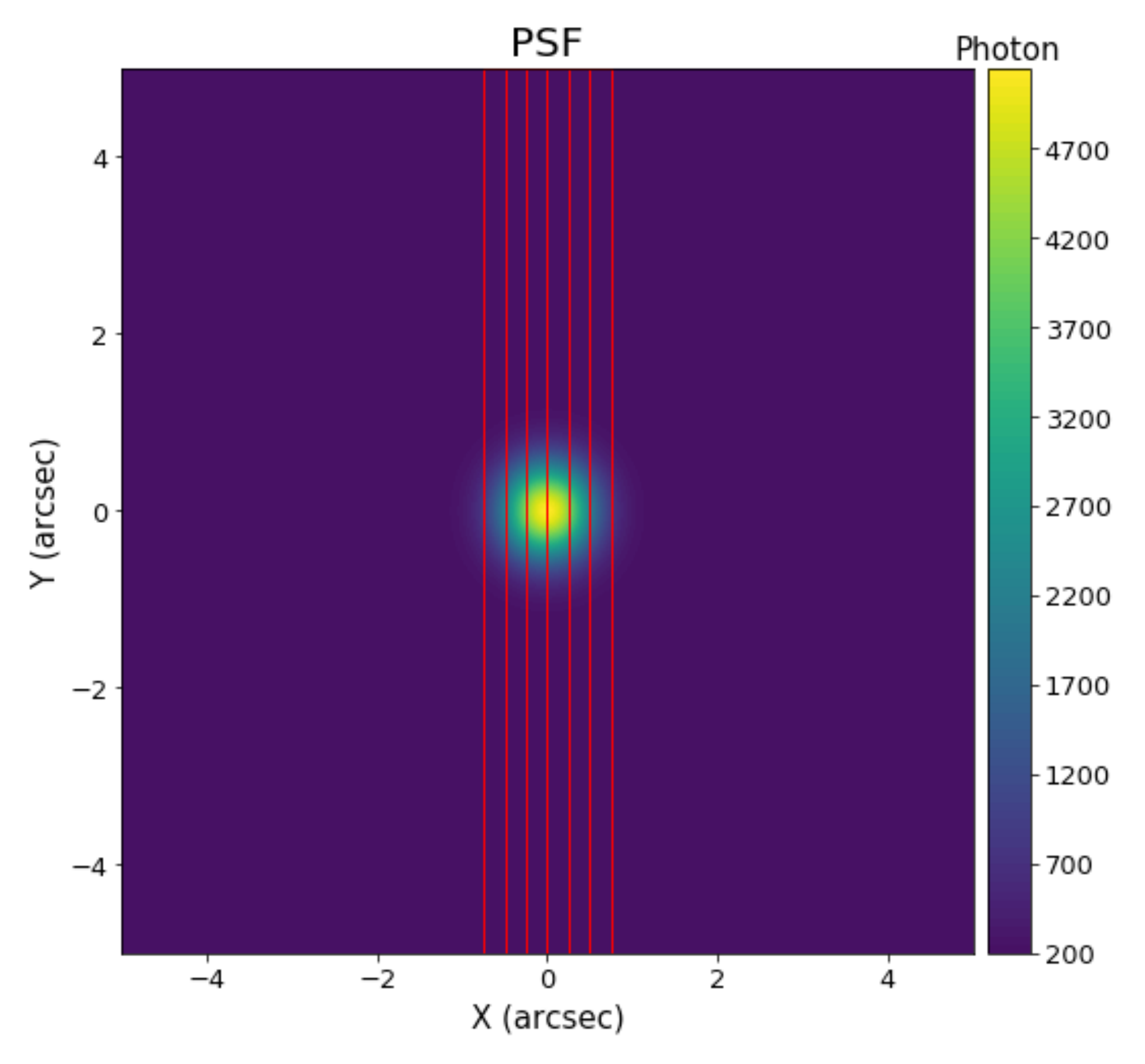}
\caption{
Simulated image of the target point spread function (PSF) on the (high-resolution) slicer focal plane, with the slice boundaries superimposed in red.}
\label{fig:E2E_Sci_02}       
\end{figure}

\paragraph{Instrument Module:} composed of the fore-optics, slicer and spectrograph units. The slicer unit computes the slit efficiency, i.e. the fraction of light passing through the pseudo-slit reformatted at the spectrograph entrance slit.  This is related to the design configuration and/or resolution mode, where the HR slicer provides an on-sky field-of-view of 
\ang{;;1.5}$\times$\ang{;;10} and the LR slicer has a larger field of 
 \ang{;;6}$\times$\ang{;;10} (as described by \cite{Ref-Zanutta_2021} and \cite{Ref-Calcines_2021}). 
The calculation of the slit efficiency is implemented taking into account the instrument entrance aperture sizes (width and length), the optical quality (PSF FWHM) of the optical path up to the slicer focal plane and the seeing (FWHM at \SI{500}{nm}). Optionally, the PSF can be convolved with a given spatial profile, which is useful to simulate the appearance of extended targets. An example of the target PSF is shown in Fig.~\ref{fig:E2E_Sci_02}.

\begin{figure}
  \includegraphics[width=0.5\textwidth]{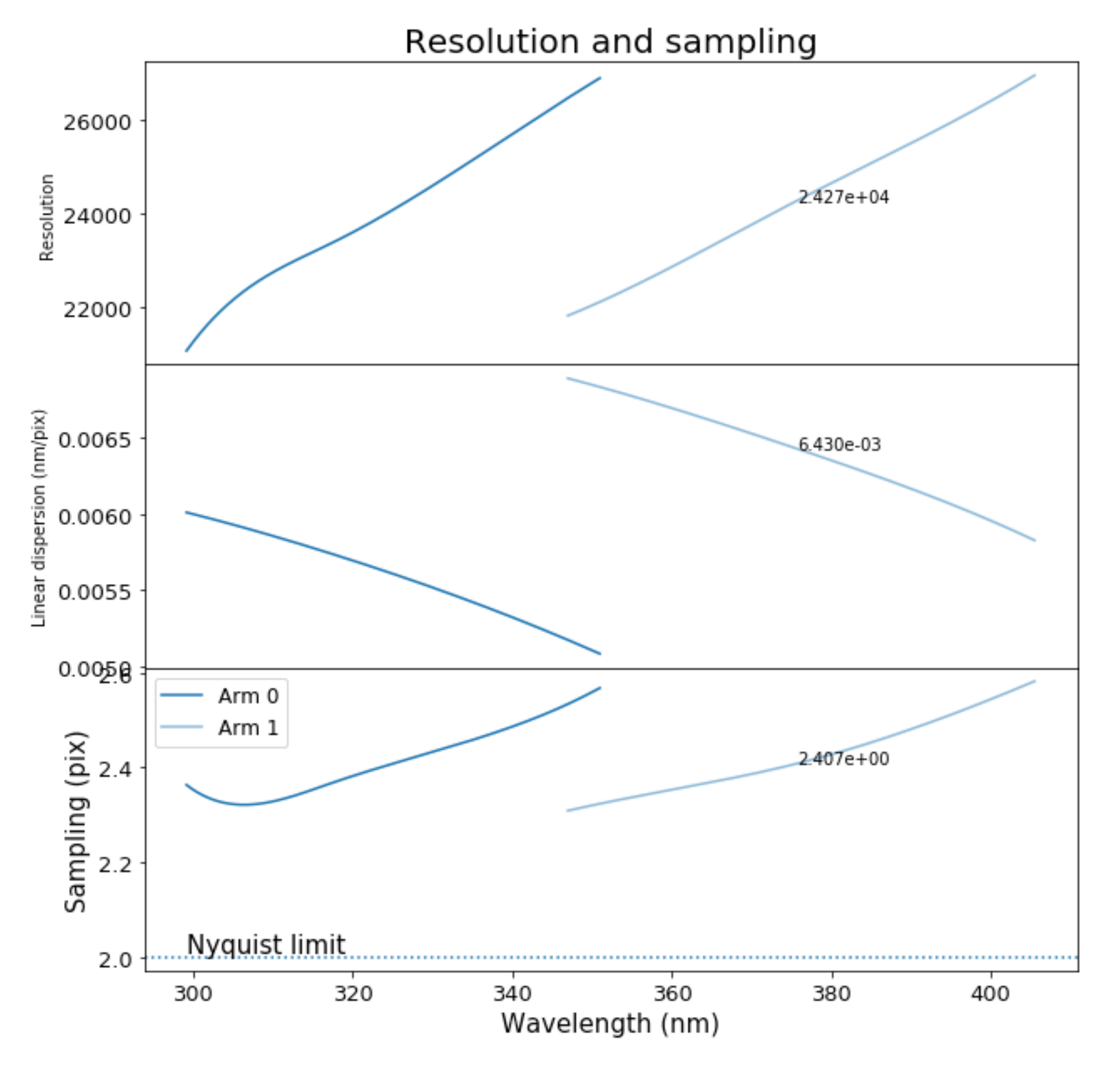}
\caption{Interpolation of resolving power, linear dispersion and sampling per resolution element in the high-resolution mode (from the static tables used by the \textit{Basic Version} of the E2E simulator). Note that the spectral sampling is above the Nyquist limit over the full 300-\SI{405}{nm} range.}
\label{fig:E2E_Sci_03_Maps}       
\end{figure}

The spectrograph unit predicts the spectral format at the focal plane of the two CUBES arms, the instrument throughput and a database of PSF maps for a set spectral resolution element (which are used by the Image Simulation Module to render the synthetic frames). 
The \textit{Parametric Version}, as explained in \cite{Ref-Genoni_2020} and \cite{Ref-Genoni_PPM}, implements the parametric equations of the different optical elements to derive the spectral format, resolving power, sampling and linear dispersion variation across the wavelength range of both arms.
Tabulated solutions of the parametric equations are also used by the \textit{Basic Version} to interpolate the sampling and linear dispersion across the simulated detector pixels and to model the spectral format (see Fig.~\ref{fig:E2E_Sci_03_Maps}). In the \textit{Full Version}, this unit extracts the slice traces (in terms of position and orientation along the main dispersion direction) and the slice PSF maps from the Spectrograph Optical File(s) designed in Zemax/OpticStudio, thus the accuracy of the aberrations, distortion and diffraction effects is defined by the parameters with which the specific optical design files tools were run.

\paragraph{Detector Module:} generates a series of maps of the effects produced by the detector, which are added to the photons distribution at pixel level by the Image Simulation Module. 

For the baseline Phase~A design with 9K-\SI{10}{\micro\meter} detectors, the expected dark current (DC) is \SI{0.5}{e^-\per pix\per\hour} (assuming a temperature of \SI{165}{K}), the adopted read-out noise (RON) is \SI{2.5}{e^-} rms (at \SI{50}{\kilo\hertz}), and a pixel-response non-uniformity of 3\% is assumed (from the manufacturer's datasheet). The detector diffusion, which spreads out photo-electrons into neighbouring pixels, is assumed to be modelled as a Gaussian kernel with a FWHM of 0.9 pixels.

\paragraph{Image Simulation Module:} renders the photon distribution for each spectral resolution element for each slice at the level of the focal plane of the spectrograph arms. 
The photons collected by each slice are mapped into separate spectral traces on the detector, applying the input template for the target and the sky to reconstruct the spectral features. In the \textit{Basic Version} this is done without consideration of any misalignment of the slices or any tilting along the main dispersion direction: all detector pixels along a given column are assumed to have the same wavelength in all the slices. 
In the \textit{Full Version}, as detailed in \cite{Ref-Genoni_2020}, this module first interpolates all of the instrumental data produced by other modules on a sub-pixel scale (of which the oversampling can be set according to the required simulation accuracy). It then produces the photon distribution for each wavelength at sub-pixel scale by convolving each slice image with the corresponding PSF map; each spectral PSF map is computed by interpolating a grid of PSF maps, extracted by the \textit{Instrument Module} from optical ray-tracing software.
The spectral slice images are properly sized and tilted according to the sampling and tilt variation along the main dispersion direction, and scaled for the integrated spectral flux and efficiency. Finally, the synthetic frame is re-binned to the pixel scale of the detectors and the noise maps from the Detector module are added.
In the current development read-out noise (RON), dark current (DC), bias levels, pixels response non-uniformity (PRNU) and pixel cross-talk (due to charge diffusion) are implemented; RON, DC and diffusion parameters are the ones reported in Table~\ref{Tab_1_Trade_Off}, while a PRNU of 4$\%$ (1$\sigma$) is assumed as educated guess at UV wavelengths.

\begin{figure*}
\centering
  \includegraphics[width=0.75\textwidth]{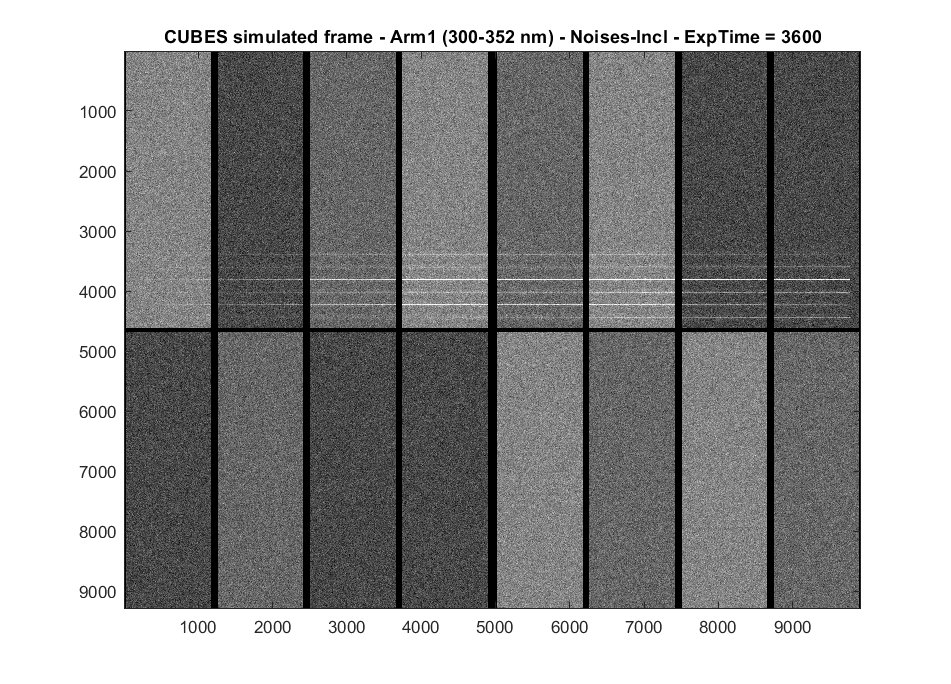}
\caption{Example simulated raw frame for Arm 1. The six slice traces are projected onto the upper half of the detector to allow the lower half to be read out separately to record the active flexure compensation (AFC) system (if analysis in the next phases shows it is required).}
\label{fig:E2E_Sci_07_Full_Frame_Arm1}       
\end{figure*}

An example of a synthetic raw frame produced by the \textit{Full Version} is shown in Fig.~\ref{fig:E2E_Sci_07_Full_Frame_Arm1}. 
The traces of the six slices can be seen, in which it is also possible to see the decreasing flux of the object towards the blue end of the spectral range (left-hand side of the figure). 
In the baseline design the traces are projected onto the upper half of the detector because the lower half is used to record the AFC fibre (if required).
In the reference condition for the HR mode the detector noise dominates over the sky background and the sky flux in the pixels cannot be easily distinguished. These frames include the pre- and over-scan regions based on the same detector model installed in other instruments (e.g. ESPRESSO on the VLT).

%

\paragraph{Extraction of the 1D spectrum:} The traces of the individual slices, as imaged by the two detectors, are extracted by a dedicated software module into a single 1D spectrum. The extraction module is currently implemented as a part of the \textit{Basic Version} of the simulator, but is actually the first simple incarnation of what will become the data reduction pipeline of the instrument.

In its current form, the extraction module simply sums the simulated pixel counts along columns, assuming that all pixels in a column have the same wavelength (which is the case for the \textit{Basic Version} implementation of the spectral format). Extraction windows are centered at the peak of the spatial profile of the simulated target in each slice and their width is defined as 1.2 times the FWHM of the profile. The extracted spectrum with its propagated error is used to estimate the SNR as a function of wavelength, given the instrument setup and adopted observing conditions. 

In its final form, the extraction module will become a part of the instrument DRS and will implement a more sophisticated extraction procedure, under the assumption that all pixels are calibrated at a different wavelength (see Section \ref{subsec_E2E_DRS}).

\section{The E2E role in the design phases}
\label{sec_E2E_app}
In this section we focus on the experimental application of the E2E simulator to evaluate the instrument design performance and functionalities while fulfilling the top-level requirements (TLRs) on the instrument, as well as  driving initial development of pipeline recipes for the DRS.

Given the scientific cases assembled for the development of CUBES, we identified a set of key performance TLRs that must be satisfied, including:
\begin{itemize}

\item CUBES shall provide a spectrum of the target over the entire wavelength range from \SIrange{305}{400}{nm} in a single exposure. (Goal: 300-\SI{420}{nm}).

\item In any part of the spectrum, the resolving power ($R$) shall be greater than 19000. The average value of $R$ shall be greater than 20000. $R$ is defined as the FWHM of unresolved spectral lines of a hollow cathode lamp in the spectral slice.

\item In a one-hour exposure the spectrograph shall be able to obtain $\textrm{SNR}=20$ for an A0-type star with $U=17.5$\,mag., for a wavelength pixel of \SI{0.007}{nm} at \SI{313}{nm} (at an airmass of 1.16). For different pixel sizes and airmasses, the SNR shall scale accordingly. 

\end{itemize}
We applied the versatility of the E2E simulator to carry out trade-off analyses of different potential designs for the HR mode, and to also assess the performance of a LR mode for the instrument.
In addition, the tool was applied by other groups to assess specific science cases (e.g. \cite{Ref-Alcala_2021}).

\subsection{Trade-Off Analysis}
\label{subsec_TradeOff}
To aid the opto-mechanical design we performed detailed trade-off analyses with the E2E \textit{Parametric Version} to evaluate the performance of different design configurations of the HR mode compared to the TLRs. We explored five different configurations, as summarized in Table~\ref{Tab_1_Trade_Off}. These considered different detector geometries, wavelength coverage and grating parameters, with the objective of assessing the instrument performance in terms of SNR and resolving power. We now briefly describe the configurations considered and our main findings. 

\paragraph{Configuration 6K:} this design used 6K-\SI{15}{\micro\meter} detectors, which were considered to minimize the spatial sampling (thus trying to maximize the SNR). The \ang{;;1.5} aperture is sampled with six slices, resulting in a \SI{0.2}{mm} input slit width to the spectrograph. 
The collimated beam size is \SI{150}{mm} and the grating parameters are well within current manufacturing capabilities. 
The geometrical spectral resolution element (SRE) sampling at the central wavelength for both arms is $\approx$2.12 pixels, while the minimum is 2 pixels (so at the Nyquist limit). The blurring term of optics PSF, flexure and detector diffusion broadens the SRE image such that the FWHM of the Gaussian fit of the resulting line spread function (LSF) at the minimum wavlength is $\approx$2.05 pixels. 
The related minimum $R$ is 18500, thus just below the requirement, while the average is 20500. 
The wavelength range provided is 305-\SI{400}{nm}, just fulfilling the minimum coverage. 
The estimated SNR for the reference object and sky conditions from the TLRs (and scaled to \SI{0.007}{nm\per pix}) satisfies the requirement (i.e. SNR\,$\ge$\,20) for both 1\,$\times$\,1 and 1\,$\times$\,2 binning (where first and second values are the spectral and spatial binning, respectively). 

\paragraph{Configurations 9K:} Adopting a 9K-\SI{10}{\micro\meter} detector and reducing the geometrical projection of the spectral resolution element allows us to extend the wavelength range to 300-\SI{405}{nm}. Within the 9K detector geometry we considered four different configurations, named H1-H2-H3-H4.

Simulations showed that increasing the grating parameters (line density and angle of incidence) while maintaining an optical PSF below one pixel enables an increase in the spectral resolving power. This is shown by the entries for configurations H1, H2 and H3 in Table~\ref{Tab_1_Trade_Off}, where the grating line-density is close to the current limit of manufacturing (for details see \cite{Ref-Gratings_2021}\footnote{A small difference in the grating parameters between the two contributions is due to the different times along the project when this analysis was performed and the final grating parameters were frozen.}) and $R_{\rm min}$\,$>$\,20000.

The differences between configurations H1 and H2 are the spectrograph input slit, where it is reduced in H2 to \SI{0.191}{mm} to help increase $R$ and the related spectral sampling. As the linear dispersion and the total number of integer pixels counted in the spatial direction for the SNR computation are the same for these two design options, the SNR values computed per pixel bin are also the same.

Configuration H3 simulates an entrance aperture of \ang{;;1.35}$\times$\ang{;;10} so the slit width and sampling are reduced by 10 $\%$ compared to H1 and the geometrical $R$ is higher by 10 $\%$. As the minimum sampling is lower than configurations H1 and H2, the blurring terms have a larger impact on the resulting LSF such that the effective gain in $R_{\rm min}$ is only 5$\%$ higher than H1 and H2. The reduced entrance aperture of \ang{;;1.35} decreases the slit efficiency, thus reducing the object counts integrated in the SNR (albeit also reducing the sky noise contribution), resulting in lower SNR values than H1 and H2.

Configuration H4 simulates the entrance aperture of \ang{;;1.5} with five slices instead of six. The collimated beam size is \SI{180}{mm}, which would translate to a significant mass increase (possibly exceeding the mass limit requirement of \SI{2.5}{ton}). Moreover, the grating parameters are more challenging with respect to other configurations. This gives $R_{\rm min}$\,$=$\,19200, so below the minimum requirement of 20000, and the $F/1.95$ camera could imply even larger optical aberrations with respect to the assumed value of 0.87 pixels FWHM of the optical PSF. The real gain of this design option is that the SNR is larger than other configurations because of the reduced number of slices.

\bigskip

From the trade-off considerations above, configuration H2 was the preferred design as it maximizes the entrance aperture with R\,$>$20000 for the whole wavelength range, while keeping the instrument within the mass limit requirement and the optics within reach of current manufacturing capabilities.
Finally, adopting this configuration, the design is able to reach a SNR $>$20 in the defined observing conditions and for the required wavelength bin. 

\begin{table*}
\caption{Trade-off results. Instrumental and performance parameters reported for the 6K-\SI{15}{\micro\meter} detector design option, and the H1-H2-H3-H4 options with 9K-\SI{10}{\micro\meter} detectors. The sampling related to the Line Spread Function (LSF) FWHM is computed as the Gaussian fit of the profile resulting from convolution of the geometrical sampling profile with the optics PSF, blurring due to flexure and detector diffusion among neighbouring pixels; all these blurring distributions are modelled as Gaussians with FWHMs as reported. The Resolving Power is that of the LSF FWHM. The SNR is computed at \SI{313}{nm}, close to the beryllium doublet, for a wavelength pixel of \SI{0.007}{nm}, and for both 1\,$\times$\,1 and 1\,$\times$\,2 (2 in spatial direction) pixel binning.
}
\label{Tab_1_Trade_Off}       
\begin{tabular}{lllllll}
\hline\noalign{\smallskip}
\textbf{\textit{Parameter}} & Units & \textbf{Config 6K} & \textbf{Config H1} & \textbf{Config H2} & \textbf{Config H3} & \textbf{Config H4} \\
\noalign{\smallskip}\hline\noalign{\smallskip}
\textbf{\textit{Instrumental}} & & & & & & \\
\noalign{\medskip}
\textit{$\lambda$ range} & \SI{}{nm} & 305-352 & 300-352 & 300-352 & 300-352 & 300-352 \\
 &  & 346-400 & 346-405 & 346-405 & 346-405 & 346-405 \\
\noalign{\smallskip}
\textit{Entrance Aperture} & \SI{}{arcsec^2} & 1.5$\times$10 & 1.5$\times$10 & 1.5$\times$10 & 1.35$\times$10 & 1.5$\times$10 \\
\noalign{\smallskip}
\textit{N. of slices} & [-] & 6 & 6 & 6 & 6 & 5 \\
\noalign{\smallskip}
\textit{Input Slit Width} & \SI{}{mm} & 0.2 & 0.2 & 0.191 & 0.18 & 0.2 \\
\noalign{\smallskip}
\textit{Beam size} & \SI{}{mm} & 150 & 154 &	160 & 154 & 180 \\
\noalign{\smallskip}
\textit{Gratings AOI} & \SI{}{deg} & 34.045 & 36.07 & 36.07 & 36.07 & 37.02\\
 & & 34.045 & 35.82 & 35.82 & 35.82 & 37.02\\
\noalign{\smallskip}
\textit{Gratings line-density} & \SI{}{l \per mm} & 3370 & 3600 & 3600 & 3600 & 3687\\
 & & 3000 & 3107 & 3107 & 3107 & 3198 \\
\noalign{\smallskip}
\textit{Camera $F/\#$} & [-] & $F/3.20$ & $F/2.60$ & $F/2.54$ & $F/2.60$ & $F/1.95$\\
\noalign{\smallskip}

\textit{RON} & \SI{}{e^-} rms & 2 & 2.5 & 2.5 & 2.5 & 2.5 \\
\noalign{\smallskip}
\textit{DC} & \SI{}{e^- \per pix \per\hour} & 3 & 0.5 & 0.5 & 0.5 & 0.5 \\
\noalign{\smallskip}
\textit{Sampling GEO at min $\lambda$} & \SI{}{pix} & 2 & 2.4 & 2.3 & 2.13 & 2.13 \\
\textit{Sampling GEO central $\lambda$} & \SI{}{pix} & 2.12 & 2.56 & 2.5 & 2.3 & 2.3 \\
\noalign{\smallskip}
\textit{Gauss FWHM Optics PSF} & \SI{}{pix} & 0.6 & 0.65 & 0.65 & 0.65 & 0.87 \\
\noalign{\smallskip}
\textit{Flexure Blur FWHM} & \SI{}{pix} & 0.5 & 0.5 & 0.5 & 0.5 & 0.5 \\
\noalign{\smallskip}
\textit{CCD Diffusion} & \SI{}{pix} & 0.9 & 0.9 & 0.9 & 0.9 & 0.9 \\
\noalign{\smallskip}
\noalign{\smallskip}\hline\noalign{\smallskip}
\textbf{\textit{Performance}} & & & & & & \\
\noalign{\medskip}
\textit{Sampling LSF FWHM at min $\lambda$} & \SI{}{pix} & 2.05 & 2.4 & 2.35 & 2.25 & 2.30 \\
\noalign{\smallskip}
\textit{Resolving Power min} & [-] & 18500 & 20500 & 21000 & 22000 & 19200 \\
\noalign{\smallskip}
\textit{SNR at 313 nm  (0.007 nm/bin) 1x1} & [-] & 25.5 & 26 & 26 & 24.5 & 26.6 \\
\noalign{\smallskip}
\textit{SNR at 313 nm (0.007 nm/bin) 1x2} & [-] & 27.5 & 28 & 28 & 27.6 & 29.2 \\
\noalign{\smallskip}\hline
\end{tabular}
\end{table*}

\subsection{Low-resolution mode}
\label{subsec_LRmode}
The investigation of possible design configurations for the LR mode of CUBES was also done using the \textit{Parametric Version} of the E2E, with similar assumptions as described for the HR trade-off above.
After some preliminary evaluations, in collaboration with the Science Team, two possible LR modes were explored:

\begin{itemize}
\item Total input aperture of \ang{;;6} and six slices (i.e. \ang{;;1} per slice). The average geometrical $R$ is then $\sim$6000 given the H2 design configuration.
\item Total input aperture of \ang{;;3} and four slices (i.e. \ang{;;0.75} per slice). The average geometrical $R$ is then $\sim$8000 given the H2 design configuration.
\end{itemize}

\begin{figure*}
  \includegraphics[width=1\textwidth]{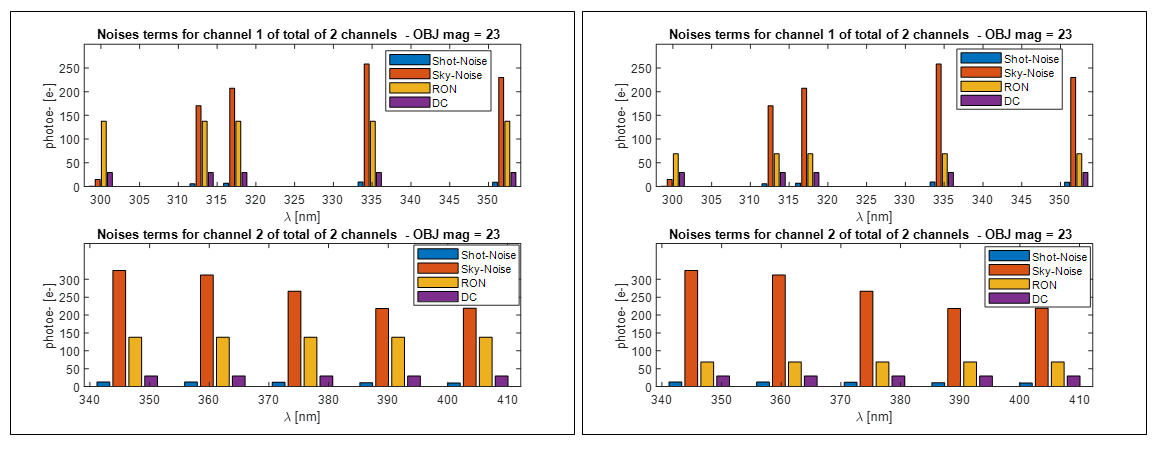}
\caption{Noise terms for the low-resolution mode with a total input aperture of \ang{;;6} and six slices. {\it Left:} 1\,$\times$\,1 binning. {\it Right:} 1\,$\times$\,2 binning.}
\label{fig:E2E_Tech_10_LR_mode_6asec}       
\end{figure*}

\begin{figure*}
  \includegraphics[width=1\textwidth]{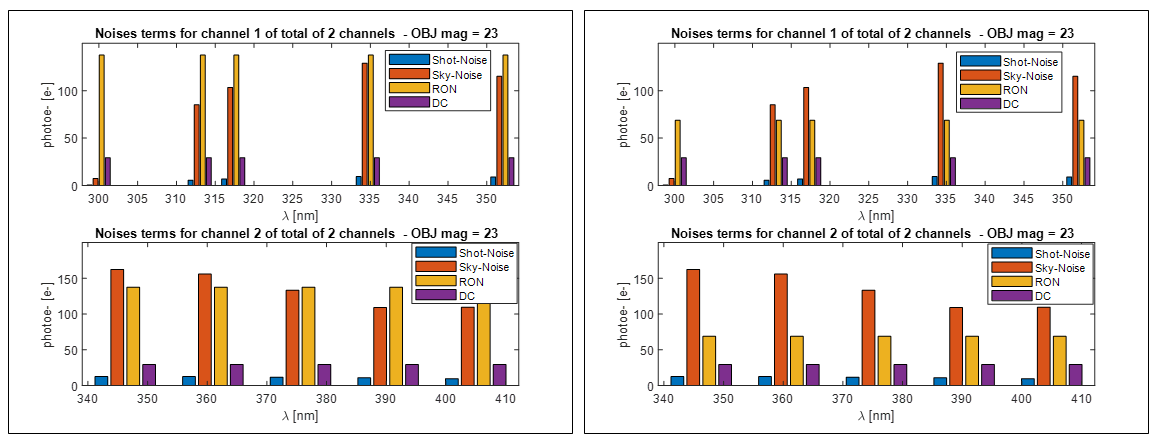}
\caption{Noise terms for the low-resolution mode with a total input aperture of \ang{;;3} and four slices. {\it Left:} 1\,$\times$\,1 binning. {\it Right:} 1\,$\times$\,2 binning.}
\label{fig:E2E_Tech_10_LR_mode_3asec}       
\end{figure*}

The noise terms were calculated using the same reference input parameters as for the HR simulations above, with the exception that only the two central slices were considered for the noise calculation (as these will capture the majority of the light in LR mode for a seeing of \ang{;;0.87}). 
The noise-terms plots for the LR mode with a total input aperture of \ang{;;6} and six slices are shown in Fig.~\ref{fig:E2E_Tech_10_LR_mode_6asec} (left-hand panel: 1\,$\times$\,1 binning, right-hand panel: 1\,$\times$\,2 binning). Similar plots are shown in Fig.~\ref{fig:E2E_Tech_10_LR_mode_3asec} for the case of a total input aperture of \ang{;;3} and four slices.

%

The conclusion from the noise-term evaluation is that the LR mode, with a \ang{;;6} aperture and six slices, will allow us to be sky-noise dominated for both detector binning options down to 313/\SI{320}{nm}. 
The other LR mode is sky-noise dominated only with 1\,$\times$\,2 binning. Other binning options (e.g., 2\,$\times$\,2) could provide further advantages, if acceptable for the scientific objectives. 

In summary, by employing slices that are four times wider for the LR slicer, the instrument remains background limited for the faintest sources, combined with delivering a spectral resolving power that is compared to X-Shooter ($R\sim6000$, see \cite{Ref-XSH}). This is a powerful combination for some science cases (where sensitivity is more critical than resolution), as outlined by e.g. \cite{Ref-Opitom}. 

\subsection{E2E and DRS connection}
\label{subsec_E2E_DRS}
The E2E will play a pivotal role in the development of the DRS, which will remove the instrumental signature from the CUBES raw spectra and calibrate them into physical units. In the software design phase, the spectral format simulated by the E2E will provide a reliable reference to prototype the algorithms responsible for extracting the spectra of the targets in an optimal way, taking into account the slice layout and the possible tilt of the lines of equal wavelength along the slices with respect to the detector pixel grid. In the coding phase, the E2E will produce simulated calibration frames alongside the science frames, to provide a complete dataset for the development and testing of the DRS recipes.  

\begin{figure*}
\begin{minipage}[c]{.99\textwidth}
\centering
\includegraphics[width=0.6\textwidth]{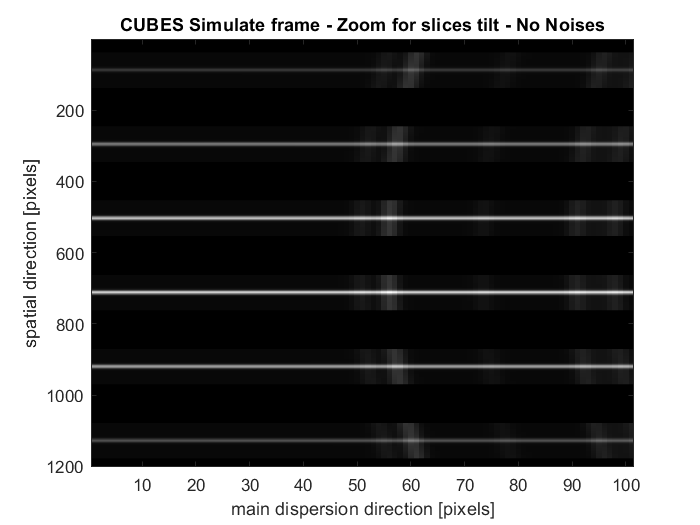}
\end{minipage}
\hspace{10mm}
\caption{Mock-up of the slice layout. Several sky lines are visible (with the most prominent located at $\sim$60 pixels in the main dispersion direction) over the continuum trace of the science target across the six slices.}
\label{fig:E2E_Tech_Tilt}  
\end{figure*}

A preliminary mock-up of the slice layout, based on the current optical configuration (H2) is shown in Fig.~\ref{fig:E2E_Tech_Tilt}. The position of the sky lines in the different slices show how the slices are misaligned with respect to each other and tilted with respect to the detector axes (especially the outermost lines). Due to these distortions, the pixels along a given detector column are not mapped to the same wavelength, and a simple summation along columns (like those described in Sect.~\ref{subsec_E2E_Modules}) is not feasible. As a consequence, the DRS will implement a more sophisticated extraction procedure, under the assumption that all pixels are calibrated at a different wavelength. The details of such an algorithm are discussed by \cite{2016SPIE.9913E..1TC}. The procedure is effectively an extension of the optimal extraction algorithm described by \cite{Horne1986}, in which the contributions of the detector pixels (each one calibrated with a specific wavelength) are also weighted by how much they overlap with the wavelength bins of the final extracted spectrum. The procedure allows creation of a single spectrum from several slices in one go, minimizing the correlation that is introduced across adjacent bins by the re-binning procedure. Furthermore, it can be naturally scaled up to combine several exposures of the same target, or scaled down to produce separate spectra of the individual slices, with a flexibility that reduction pipelines typically do not provide. In this respect, the E2E was instrumental in steering the choice of the right algorithm even before the inception of the DRS design phase.

\begin{figure*}
  \includegraphics[width=0.5\textwidth,angle=270]{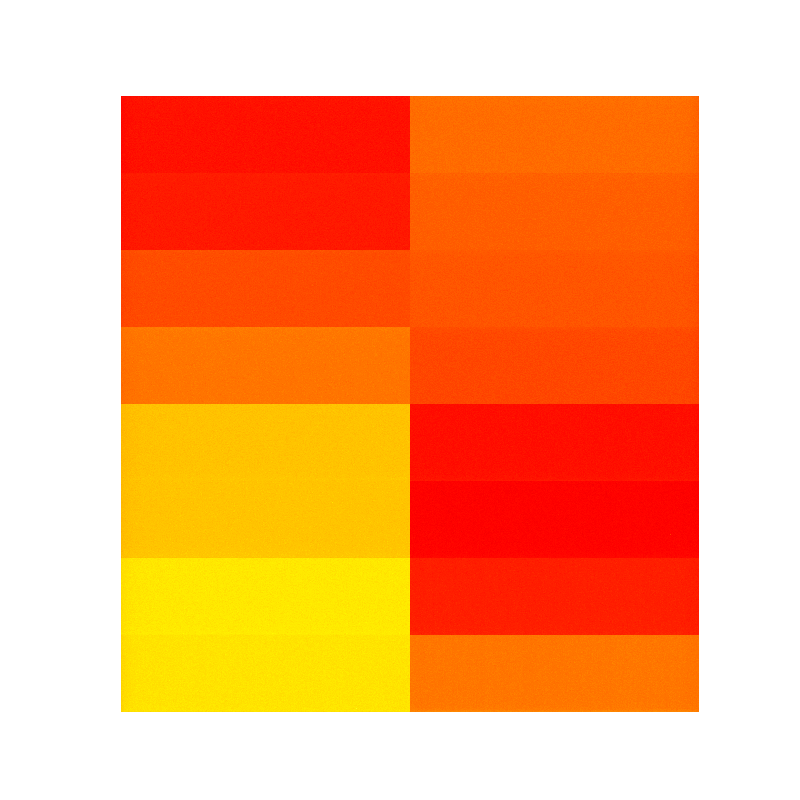}
  \includegraphics[width=0.5\textwidth,angle=270]{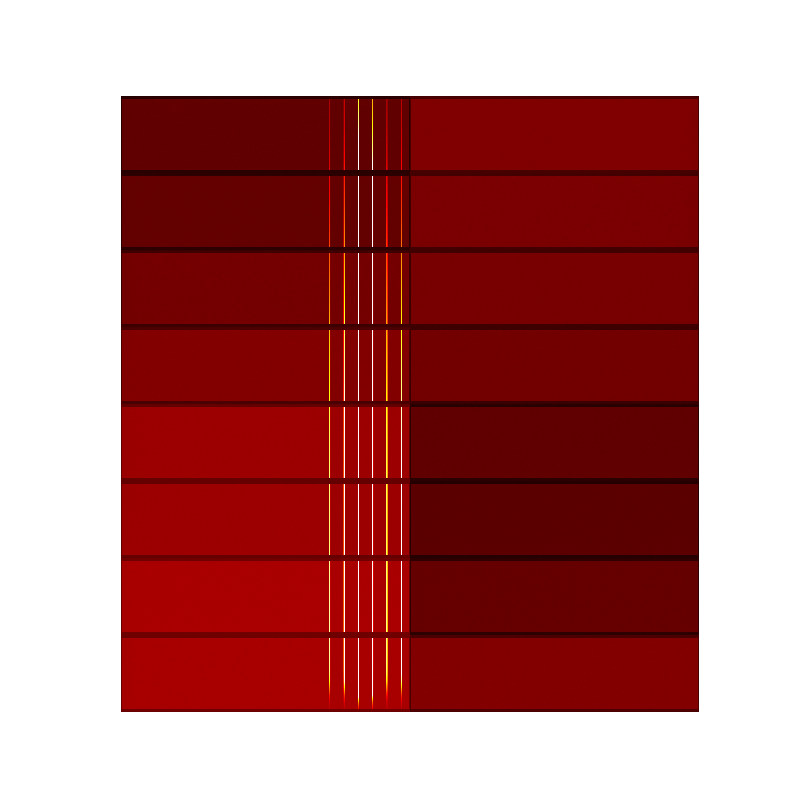}
\caption{Example of simulated calibration frames for Arm 1, constructed by similarity with VLT ESPRESSO calibration frames. {\it Left:} bias frame; {\it right:} flat frame.}
\label{fig:E2E_calibrations}       
\end{figure*}

Preliminary examples of simulated calibration frames, constructed by similarity with the calibration frames of VLT ESPRESSO, are shown in Fig.~\ref{fig:E2E_calibrations}. These frames will be integrated with other simulated calibrations and further refined throughout during the coding phase, taking advantage of the information acquired from the instrument manufacturing and integration. The interface between E2E and DRS will be settled with the definition of a shared pool of metadata to be attached to these data, and will make the two tools operate like a single entity even before the instrument comes into operation. In a sense, the E2E paradigm will be brought to its natural consequence, with the final end not being limited to a set of simulated raw data, but including their reduced counterpart, too.

\section{The CUBES ETC}
\label{sec_ETC}
Along with the E2E simulator, the CUBES ETC has provided a valuable tool during the Phase~A study to support evaluation of the science cases and the new  scientific opportunities that can be achieved with the CUBES instrument. 
During the Phase~A study we also undertook careful comparisons between the E2E and ETC for several reference input configurations, to confirm their agreement.

In general, the role of an ETC is to predict the SNR for a given exposure time. The CUBES ETC is a web-based application\footnote{http://archives.ia2.inaf.it/cubes/\#/etc} that provides the achievable SNR per pixel for a given wavelength ($\lambda$), exposure time ($t$), sky conditions, and Vega magnitude ($U$, $V$) of a point-source target. It can also provide the exposure time required to obtain a given SNR, if the SNR is given as the input instead of the exposure time.
The limiting magnitude (in AB units) to obtain SNR\,$=$\,3 at the selected wavelength and exposure time, and graphical outputs such as plots of SNR vs. $t$ and the limiting magnitude vs. SNR for the user-selected $t$, are also provided.

\subsection{ETC status and implementation}
\label{subsec_ETC_status}
\begin{figure*}
\begin{minipage}[c]{.99\textwidth}
\centering
\includegraphics[width=0.95\textwidth]{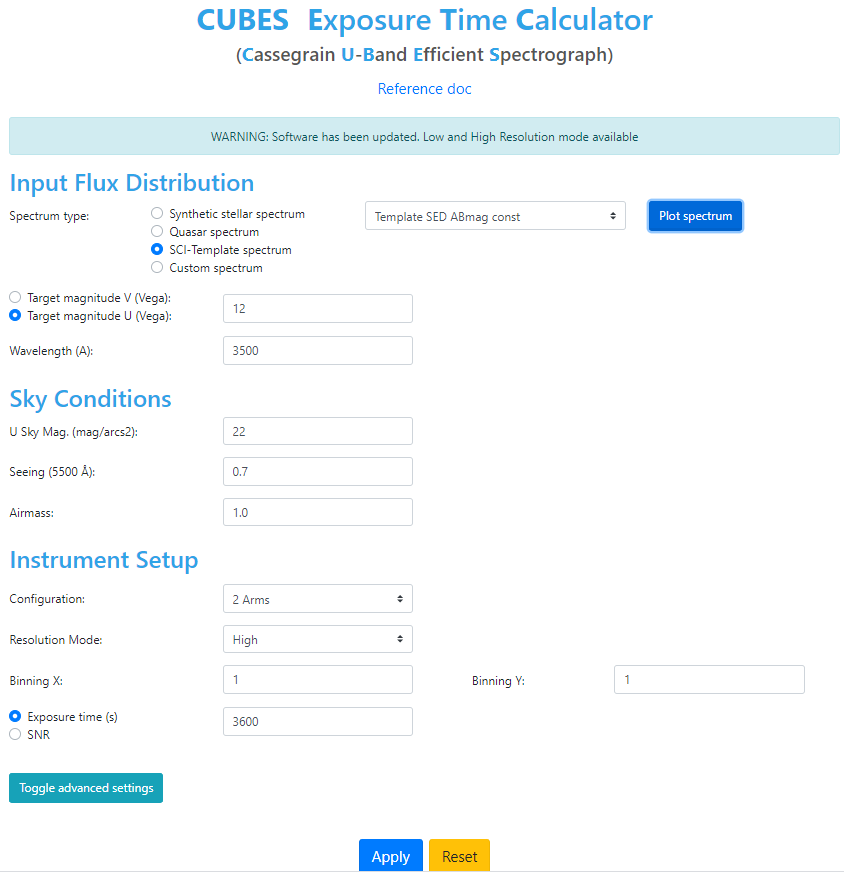}
\end{minipage}
\hspace{10mm}
\caption{Main web-page of the CUBES Exposure Time Calculator (ETC), showing the inputs required to compute the signal-to-noise ratio (SNR).}
\label{fig:ETC_MainPage}  
\end{figure*}

\begin{figure*}
\hspace{15mm}
\begin{minipage}[c]{.99\textwidth}
\includegraphics[width=0.95\textwidth]{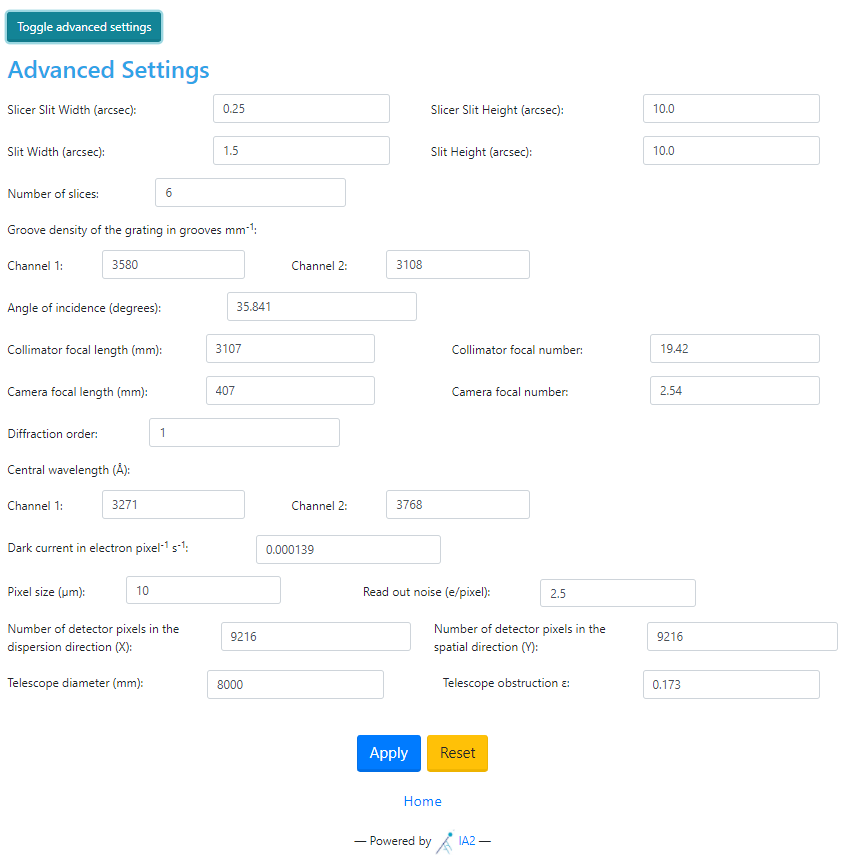}
\end{minipage}
\caption{Advanced set-up parameters in the CUBES Exposure Time Calculator (ETC) for the high-resolution mode.}
\label{fig:AdvancedSetting}  
\end{figure*}

\begin{figure*}
\begin{minipage}[c]{0.93\textwidth}
\centering
\includegraphics[width=0.8\textwidth]{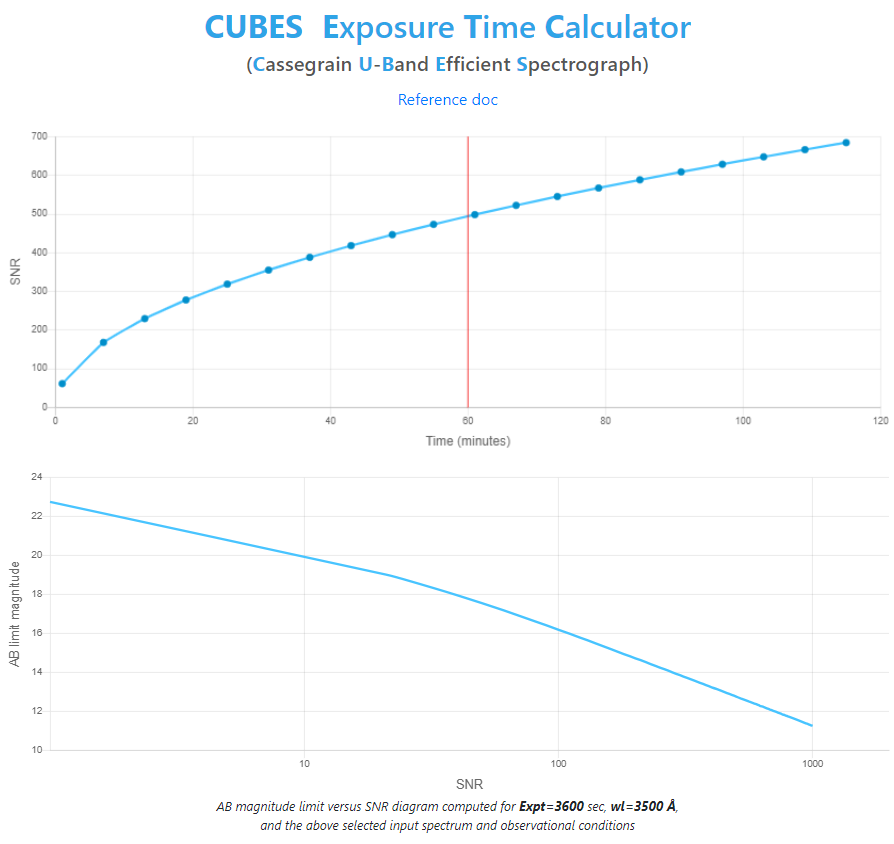}
\end{minipage}
\hspace{0mm}
\caption{Example outputs of the CUBES ETC version at the end of Phase~A. Top panel: predicted SNR vs exposure time  at $\lambda$=350\,nm for different exposure times with the HR mode of CUBES. The input template is a theoretical spectrum with constant AB magnitude and scaled to  $U$ = 12\,mag, with other parameters set to: $U_{\rm Sky}$ = 22\,mag\,arcsec$^{-2}$, airmass = 1.0, seeing = 0.87$''$,  DC = 0.5e$^-$/pix/hr, RON = 2.5\,e$^-$ rms, and 1\,$\times$\,1 pixel binning. The vertical red line shows the SNR at the selected exposure time of 1\,h. Bottom panel: SNR vs  AB magnitude limit diagram computed for the exposute time of 1\, h and the same conditions as in the top panel.}
\label{fig:ETC_outputs}  
\end{figure*}

The ETC is a Python 3.7.6 script that is called by a web application written in Java and JavaScript and uses the
facilities of the Italian Center for Astronomical Archive (IA2) operated by INAF. 
The web-page of the current ETC version consists of three parts: 
\begin{itemize}
\item The first and main part contains the observation parameters, i.e., target information, expected atmospheric
conditions, instrument configuration (LR or HR mode), and the  observation parameter exposure time or SNR  (Fig.~\ref{fig:ETC_MainPage}). A small database of input spectra (synthetic spectra, a quasar spectrum, a theoretical spectrum with constant AB magnitude) is available for use as input flux distributions. 
There is also the possibility of loading an input template (in ASCII format) from a user-deﬁned library through the option `Custom spectrum'. This input spectrum can be displayed via the `Plot spectrum' button for a quick visual check. The `Apply' button then submits the parameters to the ETC model. 

\item The second part is hidden and contains the CUBES setup, but can be visualized via the `Toggle advanced setting' at the bottom of the main one. It contains all of the instrument parameters necessary to define the HR or LR configuration of CUBES, which can be modified within their acceptable ranges. This last option allowed us to explore different design solutions during the Phase~A study. Fig.~\ref{fig:AdvancedSetting} shows the CUBES setup for the HR mode. 

\item The third part displays the computed results, which are:  graphical outputs (SNR vs. $t$ and the limiting magnitude vs. SNR as shown in Fig.~\ref{fig:ETC_outputs}), together with a table including the main input data (object type, sky condition, etc), slit losses, number of counts for the object and the sky, wavelength bin in \AA/pix, instrument efficiencies, detector binning, detector noise counts, magnitude limit at $\textrm{SNR}=3$, and SNR or $t$.

\end{itemize}

Starting from the input science spectrum (from the local database or a user-defined template), the ETC computes the SNR per spectral pixel (or spectral bin) at a user-selected $\lambda$ by taking into account the effects  produced on the target's photons by all the components encountered along their optical path from the atmosphere to the detectors. 
The ETC models of the atmosphere, telescope, slit, spectrograph, camera and detector are implemented through an appropriate parameterization. The modelling approach is similar and coherent to the that presented in the E2E description in Sect.~\ref{subsec_E2E_Modules}.
Nevertheless, we note that the purpose of the ETC is not the generation of a full synthetic frame (as in the case of the E2E simulator) but the reliable computation of SNR given a set of observing conditions, a tool for the astronomical community to explore the wide range of science cases that CUBES aims to address. 

\subsection{Example application: the Be doublet at 313\,nm}
\label{subsec_ETC_App}
The ETC has been widely used during the Phase~A study, with results presented in many of the scientific contributions in this Special Issue.
Here we highlight the example of studies of the beryllium abundance in evolved stars using the Be~{\footnotesize II} doublet at 313\,nm. 
Although one of the lightest and simplest elements, questions remain regarding the production of Be and the evolution of its abundance in the early Galaxy.

Analysis of main-sequence/turn-off stars in globular clusters with different metallicities shows that a minimum SNR of 30-40 is required to detect star-to-star variations in their Be abundances of the order of $\sim$0.2\,dex \cite{Ref-GS}.
The current limit for such studies are the 12\,hr observations with UVES of turn-off stars with $V$\,$=$\,16\,mag., where only SNR\,$=$\,8-15 was possible around the Be lines \cite{Ref-Pasquini_2004,Ref-Pasquini_2007}.

This topic has been thoroughly explored for a range of spectral types and metallicities (representative of future CUBES targets) during the Phase~A study, with detailed simulations presented by \cite{Ref-GS,Ref-Smiljanic}.
Here we simply highlight the significant performance gain provided by CUBES compared to current capabilities. For example, for a metal-poor ([Fe/H]\,$=$\,$-$2.5) input template spectrum with 
$T_{\rm eff}$\,$=$\,6500\,K and log($g$)\,$=$\,4.0, the ETC predicts SNR\,$=$\,39 for a 1\,hr exposure of a $V$\,$=$\,16\,mag. star. 

\begin{figure}
\hspace{-5mm}
\includegraphics[width=0.55\textwidth]{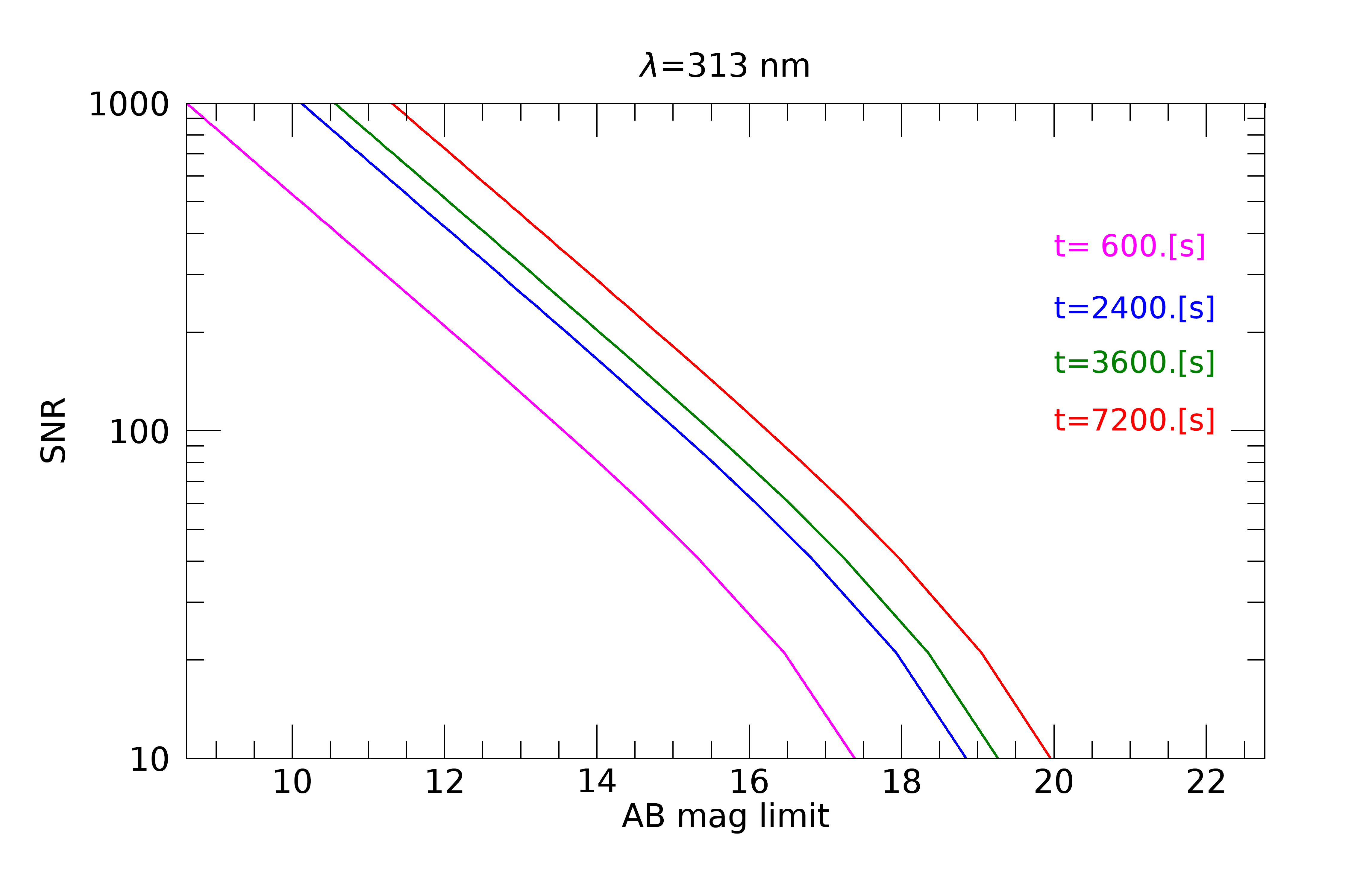}
\caption{Predicted signal-to-noise ratio (SNR) vs. AB magnitude limit at $\lambda$=313\,nm for different exposure times with the high-resolution mode of CUBES. The input template to the Exposure Time Calculator (ETC) was a theoretical spectrum with constant AB magnitude, with other parameters set to: $U_{\rm Sky}$ = 22\,mag\,arcsec$^{-2}$, airmass = 1.16, seeing = 0.87",  DC = 0.5e$^-$/pix/hr, RON = 2.5\,e$^-$ rms, and 1\,$\times$\,1 pixel binning.}
\label{fig:ETC_MAGLIM}  
\end{figure}

A further illustration of the potential performance is shown in Fig.~\ref{fig:ETC_MAGLIM}. These results show the AB magnitude limit for SNR\,$=$\,10 from the HR mode at $\lambda=\SI{313}{nm}$ for exposure times of (\SI{10}{\minute}, \SI{20}{\minute}, \SI{1}{\hour}, and \SI{2}{\hour}) for the constant AB magnitude theoretical spectrum as the ETC input and adopting: $U_\textrm{sky} = \SI{22}{mag \per arcsec\squared}$, airmass = 1.16, seeing = \ang{;;0.87}, DC = \SI{0.5}{e^-\per pix \per\hour}, RON = \SI{2.5}{e^-} rms and 1\,$\times$\,1 pixel binning. Vega and AB mags are effectively the same in the $V$-band, so with the flat input spectrum, SNR\,$>$\,40 can be recovered down to $V$\,$\sim$\,17\,mag.

\section{Conclusions}
\label{sec_Conclusions}
CUBES will provide unprecedented spectroscopic sensitivity at intermediate resolution in the bluest part of the spectrum accessible from the ground (at least from \SIrange{305}{400}{nm}). Here we have presented the package of modelling tools (E2E Simulator and ETC) developed for use by the scientific community to perform quantitative simulations of future CUBES observations and for use by the technical team in the design of the instrument.

We described the software environments as well as the modelling assumptions and approaches implemented for their development. 
We used both local environments (MATLAB and Python) and on-line web-page technologies (a Binder server interfaced through Jupyter notebook, and facilities of the Italian Center for Astronomical Archive operated by INAF).

The architecture and modelling approaches of these tools have been used to support several aspects of the Phase~A study: system design, science performance evaluation and embryonic development of the DRS. 
For instance, we have presented the trade-off analyses used to identify the most promising design configurations for the HR and LR modes. These guided important choices in terms of: entrance aperture, wavelength coverage, resolving power, SNR, manufacturing and system constraints and investigation of when the instrument is background limited compared to the expected detector noise. In the HR mode (\ang{;;1.5} aperture, six slices) the design provides $R >20000$, maximizing the entrance aperture and the wavelength coverage (300-\SI{405}{nm}), with no showstoppers in terms of system constraints. The inclusion of the second slicer for the LR mode (\ang{;;6} aperture, six slices) will provide a powerful additional capability, enabling observations to be sky-noise dominated for both detector binning options down to 313/\SI{320}{nm}.

An important first interface between simulations and the DRS has been set in the extraction of the 1D spectrum from synthetic 2D images, which is a first basic incarnation of the data reduction pipeline development; moreover, some specific aspects related to the CUBES slices tracing the spectral format and tilt of the projected slit across the detector array have been considered for the definition of future pipeline recipes.

Finally, we showed an example of how the ETC has been used to investigate the limiting magnitude of CUBES in estimating the abundances of beryllium in main-sequence stars. This example demonstrates that CUBES will be able to incresae the sample of metal-poor stars needed to address fundamental open questions pertaining to nucleosynthesis of the elements.

\section*{Acknowledgments}
The INAF authors acknowledge financial support of the Italian Ministry of Education, University, and Research with PRIN 201278X4FL and the "Progetti Premiali" funding scheme.

\section*{Conflict of interest}
The authors declare that they have no conflict of interest.

\section*{Data Availability Statement}
The paper details about simulations and it is not based on acquired data. 
The datasets generated and analysed during the current study are available from the corresponding author on reasonable request.



\end{document}